# Watchdog-LEACH: A new method based on LEACH protocol to Secure Clustered Wireless Sensor Networks

Mohammad Reza Rohbanian[1], Mohammad Rafi Kharazmi[2], Alireza Keshavarz-Haddad[3], Manije Keshtgary[4]

1. Department of Information Technology, Shiraz University, Tehran, Iran.
*Rohbanian@gmail.com*

2. Department of Information Technology, Shiraz University of Technology, Shiraz, Iran.
*kharazmi@sutech.ac.ir*

3. Department of COMMUNICATION AND ELECTRONICS, Shiraz University, Shiraz, Iran
*keshavarz@shirazu.ac.ir*

4. Department of Computer Engineering & IT, Shiraz University of Technology, Shiraz, Iran.
*keshtgari@sutech.ac.ir*

**Abstract**

Wireless sensor network comprises of small sensor nodes with limited resources. Clustered networks have been proposed in many researches to reduce the power consumption in sensor networks. LEACH is one of the most interested techniques that offer an efficient way to minimize the power consumption in sensor networks. However, due to the characteristics of restricted resources and operation in a hostile environment, WSNs are subjected to numerous threats and are vulnerable to attacks. This research proposes a solution that can be applied on LEACH to increase the level of security. In Watchdog-LEACH, some nodes are considered as watchdogs and some changes are applied on LEACH protocol for intrusion detection. Watchdog-LEACH is able to protect against a wide range of attacks and it provides security, energy efficiency and memory efficiency. The result of simulation shows that in comparison to LEACH, the energy overhead is about 2% so this method is practical and can be applied to WSNs.

**Keywords:** *LEACH, Wireless sensor networks, clustered, Watchdog-LEACH, security*

## 1. Introduction

Wireless sensor networks (WSNs) comprise of one or more base stations (BSs) that can communicate with a number of wireless sensors via a radio link. WSNs are playing an increasingly important role in a wide-range of applications. They are used for monitoring purposes, and can be used in different application areas, ranging from battlefield reconnaissance to environmental protection. Cluster-based communication protocols (e.g., [1]) have been proposed for ad hoc networks in general and sensor networks in particular for various reasons including scalability and energy efficiency. In cluster-based networks, nodes are organized into clusters, with cluster heads (CHs) that they relay messages from ordinary nodes in the cluster to the BSs. As our approach is based on LEACH protocol we describe LEACH protocol here.

LEACH [1] is a clustering-based protocol that minimizes energy dissipation in sensor networks. The purpose of LEACH is to select sensor nodes randomly as cluster heads, so the high energy dissipation in communicating with the base station is spread to all sensor nodes in the sensor network.

The operation of LEACH is separated into two phases: the set-up phase and the steady phase. The duration of the steady phase is longer than the duration of the set-up phase in order to minimize the overhead. During the set-up phase, a sensor node n chooses a random number between 0 and 1. If this random number is less than a predetermined threshold, t, the sensor node becomes a cluster head. The threshold t is calculated as:

$$t = \begin{cases} \dfrac{P}{1 - P \times [r \bmod 1/P]} & \text{if } n \in G \\ 0 & \text{otherwise} \end{cases}$$

Formula 1: Threshold to be a CH

In this equation P is the desired percentage to become a cluster head, r is the current round and G is the set of nodes that have not been selected as a cluster head in the last 1/P rounds. After a node is self-selected as a cluster head, it advertises this to all its neighbors. The sensor nodes receive advertisements and they determine the cluster that they want to belong to, based on the signal strength of the advertisements from the cluster heads. The sensor nodes inform their cluster head that they will be a member of the cluster, and then the cluster head assigns a time slot for every sensor node in which they can send data to the cluster head.

During the steady phase, the sensor nodes can begin sensing and transmitting data to the cluster heads. The cluster heads also aggregate data from the nodes in their cluster before sending them to the base station. After a certain period of time is spent in the steady phase, the network goes into the set-up phase again.

LEACH balances the draining of the energy during communication between nodes in sensor networks. The BS assumed to be directly reachable by all nodes by transmitting with high enough power. Nodes send their sensor reports to their CHs, which then combine the reports in one aggregated report and send it to the BS. To avoid the energy draining of limited sets of CHs, LEACH rotates CHs randomly among all sensors in the network in order to distribute the energy consumption among all sensors.

The recent solutions with encryption provide security against attacks, but this paper supplies CH's security and also other nodes in a new method. The rest of this paper is organized as follows: In the next section we present some related works. In Section 3, we outline security in LEACH. In Section 4, we present the Watchdog-LEACH. In Section 5, we show Energy Consumption, Finally in section 6, conclude this paper.

## 2. Related Works

Protection in the most situations is provided by encryption. For example encryption with symmetric key based security building blocks [2], or random key pre-distribution schemes [3], or using authentication framework [4]. S-LEACH [5] using two keys for each sensor node and a unique key shared with BS and the last key of the one-way key chain created by BS. S-LEACH can provide message authenticate via a message authentication code (MAC), which is computed by the unique key shared with BS. The nodes can identify the malicious CHs and choose the credible CH to join the cluster with the help of BS. However, because the choice of credible CHs over-relies on the participation of BS, the expansion of WSNs becomes impossible. Secure LEACH (Sec-LEACH) [6] uses a pre-distributed static key management scheme. In Sec-LEACH, each node is pre-distributed K keys drawn from a key pool which contains P keys. One of the major advantages provided by Sec-LEACH is that the authenticated and secure communication between CH and member nodes can be achieved without the participation of BS.

The operation of Armor-LEACH [7] combines the operations of TCCA [8] and Sec-LEACH. Time-Controlled Clustering Algorithm (TCCA) provides LEACH with more energy control results in less power consumption. Sec-LEACH provides LEACH with high level of security against many kinds of attacks. Armor-LEACH proposes a solution for sensor networks communications, where it offers high level of security with high performance.

Some papers present methods for preservation of particular attacks. In [9] is proposed a security mechanism based on LEACH routing protocol against Sybil attack. The mechanism is set to start up Sybil attack detection policy based on RSSI (Received signal strength indicator) when the cluster-heads number in WSN is over a threshold. In [10] is proposed a scheme considering cluster architecture based on LEACH protocol to build a security mechanism in a query-processing paradigm within wireless sensor network. The scheme is capable of thwarting replay attack while ensuring essential properties of security such as authentication, data integrity and data freshness. TLEACH is a WSN trust protocol [11]. TLEACH contains two main components, the Monitoring Module and the Trust Evaluation Module. Each node also maintains a Neighbor Situational Trust Table (NSTT) filled with trust value entries for each pair of node ids and situational operations. Situational operations, such as data sensing and routing, each have an individual trust value because nodes may not behave maliciously for all operations.

CSLEACH [12] (Centralized Secure Low Energy Adaptive Clustering Hierarchy) utilizes the gateway for key management, and trust management. CSLEACH builds on the LEACH algorithm by adding authentication, confidentiality, integrity, freshness and trust. Like LEACH, each sensor node is able to directly transmit to the gateway. Using a Key Distribution Center (KDC) approach, each node shares a unique private key with the gateway. CSLEACH uses single key pre-distribution to share a gateway private key that is used for broadcast authentication.

## 3. Security in LEACH

Like any wireless ad hoc network, WSNs are vulnerable to attacks [13, 14]. Besides the well-known vulnerabilities due to wireless communication and ad hoc, WSNs face additional problems, including sensor nodes being small, cheap devices that are unlikely to be made tamper-resistant or tamper-proof; and they are being left unattended once deployed in unprotected, or even hostile areas which makes them easily accessible to malicious parties. It is therefore crucial to add security to WSNs, especially those embedded in mission-critical applications.

Few papers focused on the security of WSNs even that the nature of this kind of networks may leads to low level of security which make these networks easy targets for intruders. Low Energy Adaptive Clustering Hierarchy (LEACH) provides an efficient communication protocol for WSNs to reduce power consumption and at the same time to provide some level of security. Though LEACH has several good qualities which have been widely accepted for various researches in the field of WSN, it has a hitch attached to it like any other WSN when we consider security factor. As control is distributed throughout the network of making self-organization possible, the CH nodes play an important role in the network. Hence they become a very attractive vulnerable point to the attackers.

Like most of the protocols for WSNs, LEACH is vulnerable to a number of security attacks including jamming, spoofing, replay. But because it is a cluster-based protocol, relying on their CHs for routing, attacks involving CHs are the most damaging. If an intruder manages to become a CH, it can do some attacks such as sinkhole and selective forwarding, thus disrupting the network. The intruder may also try to inject bogus sensor data into the network.

Attacks to WSNs may come from outsiders or insiders. In cryptographically protected networks, outsiders do not have credentials to show that they are members of the network, whereas; insiders do. Insiders may not always be trustworthy, as they may have been compromised, or have stolen their credentials from some legitimate node in the network.

This paper suggests a new solution to prevent WSN from being affected by intruders and thwarting some attacks.

## 4. Watchdog-LEACH

In this new method we have added some parts to LEACH protocol to make it more secure.

In Watchdog-LEACH we use *Spontaneous Watchdogs* approach [15] that we have added it before steady phase. Also we use Decentralized Intrusion Detection [16] method in setup and steady phase for Intrusion Detection. On figure 1, LEACH protocol stages and on figure 2, Watchdog-LEACH phases has been shown.

### 4.1. Watchdog-LEACH Phases

#### 4.1.1. Phase 1 _ Setup Phase

It's like LEACH setup phase as we described it in Introduction section (section1) but except first round, we have intrusion by Watchdog nodes for probable attacks too.

#### 4.1.2. Phase 2 _ Selecting Watchdog nodes

For selecting watchdog nodes we use *Spontaneous Watchdogs* approach as following.

Each node has a monitoring module that is activated upon selecting as a watchdog node. This monitoring module must be in charge of analyzing packets that their neighbors in a cluster send and receive. Maybe they receive some packets from other clusters but they ignore signals from other clusters.

Other sensors also can generate some alerts depends on the situation so their monitoring module can be started to work in some situations like attacks against the physical or logical safety of sensor but they don't check other nodes communications.

Due to the broadcast nature of communications, every watchdog node will receive all packets sent inside its cluster.

A single node will select itself as a watchdog for a cluster with a probability of 1/n. This process can resemble as *n* people with 1 dice of *m* sides each, where *n* = *m*, trying to obtain a 1 in the dice to activate the watchdog node.

If a node has been selected as a CH in previous phase, it's not selected as a watchdog in a cluster.

There isn't any extra energy consumption overhead for the decision of being a watchdog node as we assume that the result has been calculated before and has been embedded to the sensors.

**Setup phase:**

1. $H \Rightarrow g \;:\; id_H, adv$
2. $A_i \rightarrow H \;:\; id_{A_i}, id_H, join\_req$
3. $H \Rightarrow g \;:\; id_H, (\ldots, \langle id_{A_i}, T_{A_i} \rangle, \ldots), sched$

**Steady phase:**

4. $A_i \rightarrow H \;:\; id_H, D_{A_i}$
5. $H \rightarrow BS : id_H, id_{BS}, \mathcal{F}(\ldots, D_{A_i}, \ldots)$

The various symbols denote:

$A_i, H, BS :$ Sensor Nodes, CH, Base Station

$g :$ All WSN Nodes

$id_{A_i}, id_H, id_{BS} :$ ID for nodes, CHs, BS

$\rightarrow, \Rightarrow :$ Communication between nodes

$adv, join\_req, sched :$ Message kinds

$\langle id_{A_i}, T_{A_i} \rangle :$ Data message from CH to other cluster nodes including their IDs and time frame for sending data

$D_x :$ Sensed data by Node x

$\mathcal{F} :$ Aggregation Function

Figure 1: LEACH protocol

**Setup phase:**

1. $H \Rightarrow g \;:\; id_H, adv$
   $w :$ Except first roun:
   if $IsAttack(id_H, adv)$, Find $(id_H, Attack_{id})$ then Add 1 to the Sum
   if not exist, Store $(id_H, Attack_{id}, Time, Sum = 0)$

2. $A_i \rightarrow H : id_{A_i}, id_H, join_{req}, id_H \notin Blacklist$
   $w:$ Except first round:
   if $IsAttack(id_{A_i}, id_H, join_{req})$, Find $(id_{A_i}, Attack_{id})$ then Add 1 to the Sum
   if not exist, Store $(id_{A_i}, Attack_{id}, Time, Sum = 0)$

3. $H \Rightarrow g : id_H, (\ldots, \langle id_{A_i}, T_{A_i} \rangle, \ldots), sched$
   $w:$ Except first roun:
   if $IsAttack(id_H, (\ldots, \langle id_{A_i}, T_{A_i} \rangle, \ldots), sched)$, Find $(id_H, Attack_{id})$ then Add 1 to the Sum
   if not exist, Store $(id_H, Attack_{id}, Time, Sum = 0)$

**Selecting Watchdog nodes phase:**

4. $\mathcal{w}: Select\ itself\ as\ a\ watchdog, \mathcal{w} \in g, \mathcal{w} \notin H$

**Steady and Intrusion Detection phase:**

5. $A_i \rightarrow H: id_H, D_{A_i}\ (A_i \notin \mathcal{w})$
   $A_i \rightarrow BS: id_{A_i}, Attack_{id}, Time, Sum\ (A_i \in \mathcal{w},\ Sum > Threshold)$
   $\mathcal{w}: if\ IsAttack(id_H, D_{A_i}), Find\ (id_{A_i}, Attack_{id}) then\ Add\ 1\ to\ the\ Sum\ and\ update\ time$
   $if\ not\ exist, Store\ (id_{A_i}, Attack_{id}, Time, Sum = 0)$

6. $H \rightarrow BS: id_H, id_{BS}, \mathcal{F}(\ldots, D_{A_i}, \ldots)$
   $\mathcal{w}: if\ IsAttack\ (id_H, id_{BS}, \mathcal{F}(\ldots, D_{A_i}, \ldots)), Find\ (id_H, Attack_{id}) then\ Add\ 1\ to\ the\ Sum$
   $if\ not\ exist, Store\ (id_{A_i}, Attack_{id}, Time, Sum = 0)$

7. $BS \rightarrow g, (\ldots, id_{A_i}, \ldots), id_{A_i} \in Blacklist$

Symbols as previously defined, with the following additions:

*$\mathcal{w}$: Watchdog Nodes*
*Store: Storing detected attack*
*Threshold: Maximum number of attacks that can be ignored by Watchdog*
*Black List: List of intruder nodes*

Figure 2: Watchdog-LEACH protocol

The probability that $\alpha$ nodes have to activate their monitoring module at the same time is given by the following equation:

$$f(\alpha, n, m) = \frac{PR_n^{n-\alpha,\alpha} \cdot (m-1)^{n-\alpha}}{VR_{m,n}}$$

$$PR_n^{n-\alpha,\alpha} = \frac{n!}{(n-\alpha)! \cdot \alpha!}$$

$$VR_{m,n} = m^n$$

Formula 2: Watchdog selection formula

Where $PR_n^{n-\alpha,\alpha}$ is the formula of permutation with repeated elements, $VR_{m,n}$ is the formula of r-permutations with repetition, *n* is the number of nodes that could activate their monitoring module to be watchdog (i.e., the number of nodes that are going to throw a dice), and *m* is the number of nodes that are going to influence on the probability of activating the watchdog nodes, normally equal to *n* (i.e., the number of sides of every dice).
Proof of Formula 2 is available in [15] appendix.

### 4.1.3. Phase 3 _ Steady Phase and Intrusion Detection

Steady phase in Watchdog-LEACH is similar to Steady phase in LEACH protocol with some differences. Watchdog nodes doesn't sense environment signal and they work only as watchdog and monitor the cluster communications both regular nodes and CH. This behavior prevents from collisions as Watchdog-LEACH is TDMA-Based in Steady phase.

### 4.2. Intrusion Detection

Watchdog nodes are independent from CHs so they also monitor CHs behavior. During Steady phase, watchdog nodes listen to communications and when they detect an attack they send alarm to BS directly. In Watchdog-Leach we use Decentralized Intrusion Detection [16] approach for detecting attacks and sending report to BS. After each round, Black listed nodes are reported to all nodes so after that, other nodes will ignore messages from black listed nodes and they will not be selected as CH in next rounds.

The proposed algorithm was divided into the following phases:

**4.2.1. Data acquisition**: in this phase, messages are collected in a promiscuous mode and the important information is filtered before being stored, for subsequent analysis. Data extracted from the messages are stored in an array data structure and discarded after a given period of time or when there is no space left in memory.

**4.2.2. Rule application**: this is the processing phase, when the rules are applied to the stored data. If the message analysis fails the tests being applied, a failure is raised. If a message fails in one of the rules, a failure counter is incremented.

**4.2.3. Intrusion detection**: In this phase, the number of raised failures is compared to the expected amount of occasional failures in the network. If failure rate is higher than expected amount, intrusion detection is raised and an alert message is sent to BS. In other words, an attack indication is only signaled by the watchdog node when an abnormal behavior occurs with a frequency higher than expected.

On Figure 3 Intrusion Detection algorithm has been shown.

1: **for all** neighbors in Cluster **do**
2:     **for all** failure types **do**
3:         **if** round-failure value > cumulative value **then**
4:             signal attack indication
5:         **else**
6:             update cumulative value by combining it with round-failure value
7:         **end if**
8:     **end for**
9: **end for**

Figure 3: Intrusion Detection algorithm

### 4.3. Rules

**4.3.1. Interval rule:** a failure is raised if the time past between the reception of two consecutive messages is larger or smaller than the allowed limits. Two attacks that will probably be detected by this rule are the negligence attack, in which the intruder does not send data messages generated by a tampered node, and the exhaustion attack, in which the intruder increments the message sending rate in order to increase the energy consumption of its neighbors.

**4.3.2. Retransmission rule:** the monitor listens to a message, pertaining to one of its neighbors as its next hop, and expects that this node will forward the received message, which does not happen. Two types of attacks that can be detected by this rule are the black-hole and the selective forwarding attack. In both of them, the intruder suppresses some or all messages that were supposed to be retransmitted, preventing them from reaching their final destination in the network.

**4.3.3. Integrity rule:** the message payload must be the same along the path from its origin to a destination, considering that in the retransmission process there is no data fusion or aggregation by other sensor nodes. Attacks where the intruder modifies the contents of a received message can be detected by this rule.

**4.3.4. Delay rule:** the retransmission of a message by a monitor's neighbor must occur before a defined timeout. Otherwise, an attack will be detected.

**4.3.5. Repetition rule:** the same message can be retransmitted by the same neighbor only a limited number of times. This rule can detect an attack where the intruder sends the same message several times.

**4.3.6. Jamming rule:** the number of collisions associated with a message sent by the monitor must be lower than the expected number in the network. The jamming attack, where a node introduces noise into the network to disturb the communication channel, can be detected by this rule.

**4.3.7. Radio transmission range rule:** Each node must send its message with a specified power so only its neighbors can listen these messages and sending more powerful messages can be detected as hello-flood or wormhole attack.

**4.3.8 Alarm rule:** When a sensor is tampered or re-programmed or moved, a failure is raised.

**4.3.9. Intruder Watchdog rule:** When Watchdog itself is intruder, by this rule is recognized. In this case, intruder sends invalid information to BS.

**4.4. Solutions for attacks**

We have considered some attacks in our work against WSN and watchdog-LEACH's solution for them and related rule is as following.

**4.4.1. Hello Flood Attack:** (Radio transmission range rule) after Steady phase in LEACH protocol, all current clustered are canceled and new CHs start to advertise. Because watchdog nodes are cluster independent, they remain active until new clusters are formed. So watchdog nodes can listen to the advertisement messages from new CHs and if the signal strength of Hello packets are more than threshold level, so watchdog nodes send an alert to BS for quarantining the adversary node. After that the new watchdog nodes are selected by themselves.

**4.4.2. Message Delay, Repetition, Jamming**: (Interval rule, Delay rule, Repetition rule and Jamming rule) Watchdog nodes can detect it if a sensor sends messages with delay or they repeat a message. Maybe a sensor doesn't have enough energy to send messages or maybe it's under an attack so it should be excluded from network. Or maybe an adversary node sends repetitive messages to jam CH or for sending wrong information to CH.

**4.4.3. Black-hole and Selective Forwarding:** (Delay rule and Retransmission rule**)** Sometimes intruder node as a regular node or CH, doesn't send messages to related CH or BS and it drops the message or sends it to another malicious node or network. Watchdog nodes can detects these types of attacks by checking the packet destination and CH activities and communications.

**4.4.4. Physical attack:** (Alarm rule) Sometimes a sensor is damaged physically or re-programmed or moved to another place. In this case, sensor sends an alarm and watchdog nodes report it to BS.

**4.4.5. Intruder Watchdog:** (Intruder Watchdog rule) In this case, watchdog itself is an intruder and sends wrong info to the BS and reports invalid attacks. For solving this problem, other watchdogs in the cluster, compare the reported attack with their info. If there is a difference between their info and received report, they report an attack to BS.

Other attacks can be prevented with authentication and encryption or other methods that can be applied to Watchdog-LEACH protocol. We assume that we have encrypted all messages with a pre-distributed and hardware embedded key and MAC is calculated and added to each message.

### 4.5. Security Comparison with other protocols

On Table 1, we have a comparison between Watchdog-LEACH and other famous methods that have been applied on LEACH for more security.

|  | LEACH | TLEACH | CSLEACH | Watchdog LEACH |
|---|---|---|---|---|
| Integrity | No | No | MAC | Yes |
| Authentication | No | No | Pre-distributed Keys | Yes |
| Confidentiality | No | No | Symmetric Key Encryption | Yes |
| Trust | No | NSTT / Monitoring Neighbors | TC | Watchdogs |

Table 1: Comparison between Watchdog-LEACH and other famous related protocols

## 5. Energy Consumption

In a simulation we used [1] assumptions as following:

$E_{elec}$ = 50 nJ/bit

$\varepsilon_{amp}$ = 100 pJ/bit/m2 = 0.1 nJ/bit/m2

Data rate = 2000bits/s

Data package size = 2000-bit

Signal package size = 64-bit

So required values for energy consumption is calculated on table 2.

| Energy Consumption type | Value |
|---|---|
| Receive a data message | 100 µJ/message |
| Receive a signal message | 3 µJ/message |
| Send a data message (d<= 60m) | 820 µJ/message |
| Send a signal message (d<=60m) | 26 µJ/message |
| Send a message (d > 60m) | ETx_data = Eelec* k-bit/message + εamp*k*d² =100 µJ + 0.1 nJ*d² |

Table 2: Required amounts for WSN energy consumption calculation

Also we considered a WSN with 1000 nodes and 100 clusters. Total rounds is 1000 and steady phase rounds number is 10. Also we assume that maximum attacks have been occurred (10 attacks). The average distance from any node to BS is 100m. For total network energy consumption calculation, we need to calculate CH energy (Formula 3), Watchdog energy (Formula 4), Sensor energy (Formula 5) . Total Energy formula is shown on Formula 6.

$$E(CH\ in\ Setup) = E(SSM) + (n-1).E(RSM) + E(SDM)$$
$$E(CH\ nodes\ Selecting) \approx 0$$
$$E(CH\ in\ Steady\ phase) = (n-1).RDM + SDMtoBS$$
$$E(CH\ Total) = E(CH\ in\ Setup) + E(CH\ nodes\ Selecting) + E(CH\ in\ Steady\ phase).ncs$$

The various symbols denote:
$E:Energy$
$SSM:Sending\ Signal\ Message$
$SDM:Sending\ Data\ Message$
$RSM:Receiving\ Signal\ Message$
$RDM:Receiving\ Data\ Message$
$SDMtoBS:Sending\ Data\ Message\ to\ BS$
$n:Number\ of\ sensors\ in\ a\ cluster$
$ncs:Number\ of\ cycles\ in\ Steady\ phase$

Formula 3: CH energy

$$E(W\ Setup) = RSM + (n-1).RSM + RDM + SSM$$
$$E(W\ nodes\ selecting) \approx 0$$
$$E(W\ Steady\ phase\ without\ Attack) = (n-1).RDM$$
$$E(W\ Steady\ phase\ with\ Attack) = (n-1).RDM + SDMtoBS$$
$$E(W\ Total) = E(W\ Setup) + E(Selecting\ Watchdog\ node) + (ncs - noa).E(W\ Steady\ phase\ without\ Attack) + noa.E(W\ Steady\ phase\ with\ Attack)$$

Symbols as previously defined, with the following additions:

$W:Watchdog\ node$
$noa:Number\ of\ Attacks$

Formula 4: Watchdog energy

$$E(S\ Setup) = RSM + SSM + RDM$$
$$E(S\ Steady\ phase) = SDM$$
$$E(S\ Total) = E(S\ Setup) + ncs.E(S\ Steady\ phase)$$

Symbols as previously defined, with the following additions:

$S:Sensor$

Formula 5: Sensor energy

$$E(Network\ Total) = [E(CH\ Total) + (nwnc.E(W\ Total) + (ncs-nwnc-1).E(S\ Total)].noc.nosc + nons.RDM$$

Symbols as previously defined, with the following additions:

$nwnc: Number\ of\ Watchdog\ Nodes\ in\ a\ Cluster$
$noc: Number\ of\ Clusters$
$nosc: Number\ of\ sensors\ in\ a\ Cluster$
$nons: Number\ of\ Network\ Sensors$

Formula 6: Total network energy

With our assumptions we have following values (Table3) so we can calculate Total network energy consumption with consideration of Watchdog numbers in each cluster (Figure 4). According to Figure 4, with each additional Watchdog nodes, 2% is added to total energy consumption. So our protocol has about 2% energy overhead in comparison to LEACH protocol. This result shows that our proposed method is practical and can be applied on WSNs.

| Energy Formula | Value(J) |
|---|---|
| $E(CH\ in\ Setup)$ | 0.000873 |
| $E(CH\ nodes\ Selecting)$ | 0 |
| $E(CH\ in\ Steady\ phase)$ | 0.001001 |
| $E(CH\ Total)$ | 0.010883 |
| $E(W\ Setup)$ | 0.000156 |
| $E(W\ nodes\ selecting)$ | 0 |
| $E(W\ Steady\ phase\ without\ Attack)$ | 0.0009 |
| $E(W\ Steady\ phase\ with\ Attack)$ | 0.001001 |
| $E(W\ Total)$ | 0.010166 |
| $E(S\ Setup)$ | 0.000129 |
| $E(S\ Steady\ phase)$ | 0.00082 |
| $E(S\ Total)$ | 0.008329 |

Table 3: Energy Amounts

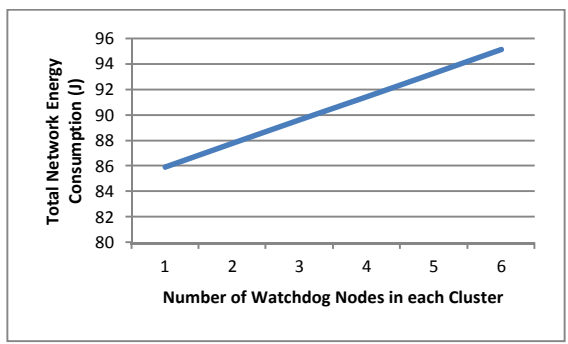

Figure 4: Total Network Energy Consumption (J) vs. Number of Watchdog nodes in each cluster

Also in Figure 5, watchdog node Energy consumption based on number of attacks has been shown. According to Figure 5 results and Table 3, in maximum number of attacks, consumed energy for a watchdog node is less than a CH. For this calculation we considered the threshold to be 1 so any time a watchdog recognizes an attack, it reports to BS. But in real protocol, it reports to BS after some attacks so the real energy consumption will be less than these results.

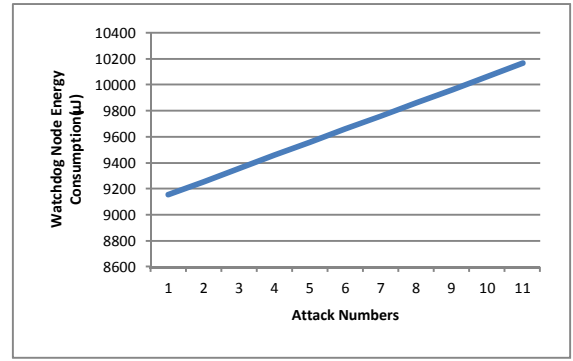

Figure 5: Watchdog Node Energy Consumption ($\mu J$) vs. Number of Attacks

## 6. Conclusion

Watchdog-LEACH can be used to increase WSN security. Also it can be combined with other methods (authentication, encryption and etc.) to provide more security for WSN. This approach is CH and cluster independent and it can detect malicious CHs too that we didn't have this feature in previous methods. Our detection is decentralized since the monitor modules are distributed on network, installed in common nodes.

Also this method in comparison to LEACH has about 2% energy overhead so this method is practical and can be applied to WSNs.

For future works, this method can be upgraded to a lightweight detection framework for prevention and detection of common attacks in WSNs.